# ANALISIS EKSPLORATIF DAN AUGMENTASI DATA NSL-KDD MENGGUNAKAN *DEEP GENERATIVE ADVERSARIAL NETWORKS* UNTUK MENINGKATKAN PERFORMA ALGORITMA *EXTREME GRADIENT BOOSTING* DALAM KLASIFIKASI JENIS SERANGAN SIBER

**Kevin Putra Santoso[1]**
*Departemen Teknologi Informasi*
Institut Teknologi Sepuluh Nopember Surabaya
5027211030@student.its.ac.id

**Fadhl Akmal Madany[2]**
*Departemen Teknik Informatika*
Institut Teknologi Sepuluh Nopember Surabaya
5025221028@student.its.ac.id

**Hatma Suryotrisongko[3]**
*Departemen Teknologi Informasi*
Institut Teknologi Sepuluh Nopember Surabaya
hatma@is.its.ac.id

**Abstrak:** Di tengah eskalasi serangan siber yang meningkat dalam kompleksitas dan frekuensi, keamanan informasi menghadapi tantangan yang belum pernah terjadi sebelumnya. Perkembangan teknologi dan ekspansi digitalisasi secara global telah memperluas vektor serangan, meningkatkan kerentanan terhadap serangan siber. Cybersecurity Ventures memprediksi bahwa biaya kejahatan siber global akan meningkat secara dramatis, sementara *World Economic Forum* menempatkan kejahatan siber sebagai salah satu risiko global teratas. Hal ini diperparah oleh fakta bahwa kurang dari 25 persen kejahatan siber dilaporkan kepada penegak hukum. Dataset NSL-KDD, sebagai standar benchmark dalam deteksi intrusi siber, menghadapi tantangan dalam klasifikasi jenis serangan, terutama karena isu ketidakseimbangan data. Metode augmentasi data tradisional seperti *oversampling* dan *undersampling* memiliki keterbatasan dalam mengatasi masalah ini. Oleh karena itu, penelitian ini mengusulkan penerapan *Deep Generative Adversarial Networks* (GANs) untuk augmentasi dataset NSL-KDD. Tujuan utama adalah untuk meningkatkan efikasi *eXtreme Gradient Boosting* (XGBoost) dalam klasifikasi serangan siber. Sebagai hasilnya, metode yang diusulkan penelitian ini memberikan akurasi sebesar **99.53%** dengan model XGBoost tanpa augmentasi data dengan GAN, dan **99.78%** dengan augmentasi data menggunakan GAN.

**Kata-kata Kunci:** Analisis Eksploratif, Augmentasi Data, NSL-KDD, *Generative Adversarial Network*, XGBoost, Serangan Siber

## 1. Pendahuluan

Dalam ranah keamanan informasi, serangan siber menunjukkan eskalasi baik dalam kompleksitas maupun frekuensi, menimbulkan ancaman krusial terhadap keamanan informasi. Perkembangan teknologi dan ekspansi digitalisasi secara global meningkatkan vektor serangan, menyebabkan kerentanan yang lebih luas terhadap serangan siber. Cybersecurity Ventures memprediksi biaya kejahatan siber global akan meningkat sebesar 15 persen per tahun, mencapai $8 triliun USD pada tahun ini dan $10.5 triliun USD pada 2025 [1]. Kejahatan siber kini menjadi salah satu dari 10 risiko global teratas menurut World Economic Forum [1]. Lebih lanjut, Cybersecurity Ventures menginformasikan bahwa kurang dari 25 persen kejahatan siber dilaporkan kepada penegak hukum pada 2023. IBM juga melaporkan bahwa biaya rata-rata dari pelanggaran data adalah $4.35 juta USD pada 2022, meningkat 2.6 persen dari 2021.

Dataset NSL-KDD, yang diakui sebagai standar benchmark dalam deteksi intrusi siber, menyajikan kompleksitas dalam klasifikasi jenis serangan [2]. Analisis oleh Abdelkhalek [3] menunjukkan bahwa dataset ini mencakup spektrum serangan yang luas namun dihambat oleh isu ketidakseimbangan data dan tidak merepresentasikan pola serangan terkini secara adekuat. Hal ini dapat menyebabkan gangguan yang signifikan pada performa model.

Beberapa statistik seperti *oversampling* dan *undersampling* dapat diterapkan dalam augmentasi data untuk menangani masalah ketidakseimbangan data. Akan tetapi, metode *undersampling* dapat mengurangi jumlah data secara signifikan apabila terdapat ketidakseimbangan yang tinggi, sedangkan penerapan teknik *oversampling* seperti SMOTE [4] dapat menyebabkan sebuah model rentan mengalami *overfitting* karena duplikasi dari data minoritas. Untuk menangani hal tersebut, Tanaka mengatakan bahwa *Generative Adversarial Networks* dapat diimplementasikan untuk membuat data baru dengan mempelajari fitur-fitur dari kelas yang ada dalam dataset [5].

Penelitian ini menerapkan Deep Generative Adversarial Networks untuk augmentasi dataset NSL-KDD, disertai dengan analisis eksploratif, untuk meningkatkan efikasi Extreme Gradient Boosting dalam klasifikasi serangan siber. Integrasi metodologi ini diharapkan mengatasi limitasi yang inheren dalam dataset NSL-KDD, sekaligus meningkatkan kapasitas deteksi serangan siber. Pendekatan ini berpotensi memberikan kontribusi substantif dalam domain keamanan siber, terutama dalam pengembangan sistem deteksi intrusi yang adaptif dan presisi tinggi.

## 2. Landasan Teori

### 2.1. *Sampling*

Pengambilan sampel (*Sampling*) adalah teknik yang melibatkan pemilihan subset S dari populasi P untuk menyimpulkan sifat-sifat P. Teknik sampel yang baik dapat diukur dari seberapa beragam sampel tersebut dan seberapa baik sampel tersebut mewakili seluruh populasi [6].

### 2.2. Kardinalitas

Kardinalitas (*Cardinality*) berkaitan dengan kekhasan nilai data dalam suatu kolom. Secara matematis jika A dilambangkan dengan $|A|$. Kardinalitas tinggi menyiratkan himpunan yang lebih besar dengan nilai yang lebih unik [7].

### 2.3. *eXtreme Gradient Boosting* (XGBoost)

XGBoost adalah model *boosted tree* yang memanfaatkan fungsi tujuan J yang dioptimalkan menggunakan penurunan gradien sebagai berikut.

$$J(\theta) = \sum_i l(y^{(i)}, \hat{y}^{(i)}) + \sum_k \Omega(f_k) \quad (1)$$

Di mana $l$ adalah fungsi kerugian pelatihan yang dapat dibedakan dan $\Omega$ adalah istilah regularisasi. Kerugian pelatihan mengukur seberapa prediktif model kita terhadap data pelatihan [8].

### 2.4. *Isolation Forest*

*Isolation Forest* mengidentifikasi anomali dengan mengisolasi *outlier* dalam langkah yang lebih sedikit daripada *inlier*. Algoritma ini bekerja dengan memilih sebuah fitur dan kemudian secara acak memilih nilai pemisahan antara nilai maksimum dan minimum dari fitur yang dipilih. Panjang jalur $h(x)$ digunakan untuk menghitung skor anomali yang didefinisikan sebagai

$$s(x, n) = 2^{-\frac{h(x)}{c(n)}} \quad (2)$$

Di mana $c(n)$ adalah panjang jalur rata-rata dari pencarian yang gagal dalam *Binary Search Tree* [9].

### 2.5. *Generative Adversarial Networks* (GANs)

GAN adalah arsitektur pembelajaran mendalam generatif yang biasa digunakan untuk menghasilkan data baru. GAN terdiri dari dua jaringan, generator dan diskriminator, yang dilatih secara bersamaan melalui proses permusuhan. Generator menghasilkan data, sedangkan diskriminator menilai apakah data tersebut terlihat asli atau palsu. Proses pelatihan melibatkan kedua jaringan untuk meningkatkan kinerjanya hingga generator menghasilkan data berkualitas tinggi, dan diskriminator hampir tidak dapat membedakan apakah suatu data itu nyata atau data yang dihasilkan. Secara matematis, GAN melibatkan generator G dan diskriminator D, dilatih melalui permainan min-max, dijelaskan dengan fungsi nilai V yang dilambangkan seperti di bawah ini.

$$\min_G \max_D V(D, G) = \mathbb{E}[\log D(\vec{x})] + \mathbb{E}[\log(1 - D(G(\vec{z})))] \quad (3)$$

Tujuan GAN adalah meminimalkan *loss function* generator dan memaksimalkan *loss function*

diskriminator. di mana $\mu_B$ dan $\sigma_{B^2}$ masing-masing adalah mean dan varians dari mini-batch B [10].

### 2.6. *Accuracy Score*
Skor akurasi digunakan untuk mengevaluasi kinerja model dalam prediksi. Dihitung sebagai:

$$Akurasi = \frac{jumlah\ prediksi\ benar}{jumlah\ total\ prediksi} \quad (4)$$

Pada umumnya, skor akurasi digunakan ketika suatu kesimpulan tidak terpengaruh oleh adanya positif palsu atau negatif palsu.

## 3. Penelitian Terkait
Hadi, dkk. (2022) merancang sistem deteksi intrusi yang mengklasifikasikan jenis serangan berdasarkan dataset NSL-KDD. Hadi menggunakan pendekatan *deep learning* dengan lima variasi model berbeda dan membandingkan performanya satu sama lain. Lima model tersebut di antaranya *deep neural network* (DNN), *convolutional neural network* (CNN), *recurrent neural network* (RNN), *long-short term memory* (LSTM), dan *gated recurrent unit* (GRU). Berdasarkan hasil penelitiannya, Hadi memperoleh nilai akurasi sebesar 98.53% untuk DNN, 98.63% untuk CNN, 98.13% untuk RNN, 98.04% untuk LSTM, dan 97.78% untuk GRU [11].

Pendekatan yang dilakukan oleh Soleymanzadeh, dkk. (2022) menggunakan *generative adversarial networks* (GAN) dan *ensemble convolutional neural networks* (E-CNN) untuk melakukan augmentasi data dan menggunakannya untuk melatih model klasifikasi. Sebagai hasilnya, model yang diusulkan oleh Soleymanzadeh memperoleh nilai akurasi sebesar 86.36% dan *F1-score* sebesar 85.81% [12].

Shone, dkk. (2018) menerapkan pendekatan semi-supervised learning untuk melakukan klasifikasi serangan berdasarkan dataset NSL-KDD. Dalam penelitiannya, Shone menerapkan model 2 tumpukan auto encoder (dengan 3 lapisan tersembunyi di dalamnya, masing-masing berjumlah 14, 28, 28 neuron) diikuti dengan *random forest classifier*. Sebagai hasil, Shone berhasil memperoleh akurasi sebesar 85.42% [13].

## 4. Analisis Data
### 4.1. *Data Preprocessing*
Pada tahap awal preprocessing data, baris yang terdiri dari *missing value* atau *NaN* akan kami hilangkan karena banyaknya data yang hilang hanya sekitar 0.1% dan untuk memastikan integritas data tetap terjaga. Nilai data yang tidak konvensional, seperti '*' dan '99999', diganti dengan 0 untuk mengelola *outlier* dan mempertahankan konsistensi dataset. *One-hot-encoding* diterapkan pada fitur 'protocol type' untuk mengakomodasi kardinalitasnya yang rendah, sementara kolom 'service' dan '*flag*' dikodekan dengan *label encoder* karena kardinalitasnya yang tinggi, sehingga mengoptimalkan dataset untuk analisis dan pelatihan model selanjutnya.

### 4.2. Analisis Distribusi Persebaran Kelas
Setelah melakukan preprocessing data, kami melakukan analisis distribusi untuk setiap kelas. Ditemukan bahwa kelas target didominasi oleh jenis interaksi normal, diikuti oleh jenis serangan 'neptune' yang sering terjadi, dan kemudian oleh 'satan' dan 'ipsweep', yang masing-masing menempati posisi kedua dan ketiga dalam kategori serangan yang paling sering terjadi.

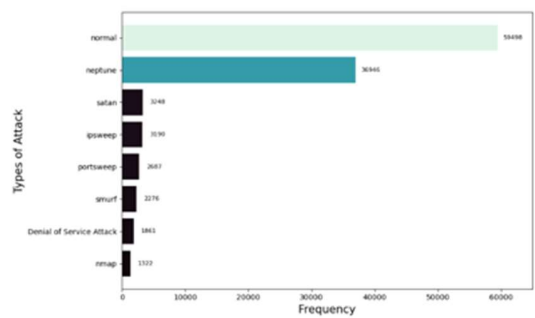

**Gambar 2.1** Distribusi Persebaran Jenis Serangan Masing-masing Label

Berdasarkan gambar diatas, dapat diamati bahwa terdapat ketidakseimbangan data yang signifikan. Hal ini akan mempengaruhi performa model dalam memprediksi kelas dengan jumlah terkecil.

### 4.3. Analisis Prediktif
XGBoost digunakan untuk model prediksi kelas serangan. *Hyperparameter Tuning* diimplementasikan setelah melatih model sebanyak 160 iterasi. Dataset dibagi dengan rasio 60:20:20 untuk set latih, set validasi, dan set uji terlebih dahulu. Melalui proses ini, model mencapai tingkat akurasi 99,54%. Konstanta untuk kepentingan fitur yang diperoleh dari model disajikan pada gambar berikut.

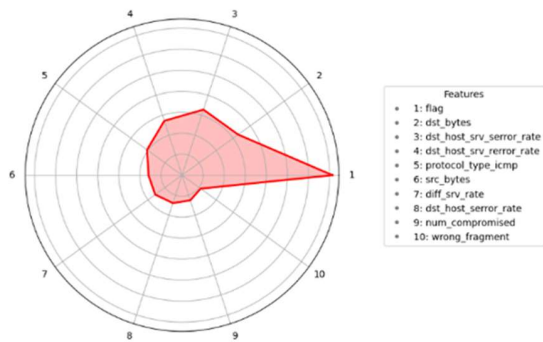

**Gambar 2.2** Angka *Feature Importances* dari Setiap Fitur

Nilai-nilai *feature importance* diperoleh berdasarkan rata-rata *Gain* dari setiap cabang pohon yang dihasilkan oleh XGBoost, dan 10 fitur teratas dengan nilai gain tertinggi ditampilkan. Gain dihitung ketika sebuah node terpecah, yang mewakili peningkatan akurasi yang dibawa oleh sebuah fitur pada cabang terkait. Ditemukan bahwa fitur 'flag', 'dst_bytes', dan 'dst_host_srv_serror_rate' merupakan tiga fitur teratas yang memiliki pengaruh signifikan terhadap klasifikasi, diikuti oleh 7 fitur lainnya seperti yang ditampilkan pada gambar.

### 4.4. Analisis Anomali

Analisis anomali dilakukan untuk mengidentifikasi jenis serangan yang tidak biasa atau jarang terjadi yang mungkin merupakan tanda dari jenis serangan baru atau tidak biasa. Untuk mendapatkan nilai anomali, *Isolation Forest* digunakan. Berdasarkan gambar 2.3, dapat dilihat bahwa serangan DDoS, smurf, dan nmap memiliki tingkat anomali yang tinggi berdasarkan karakteristik fitur-fiturnya.

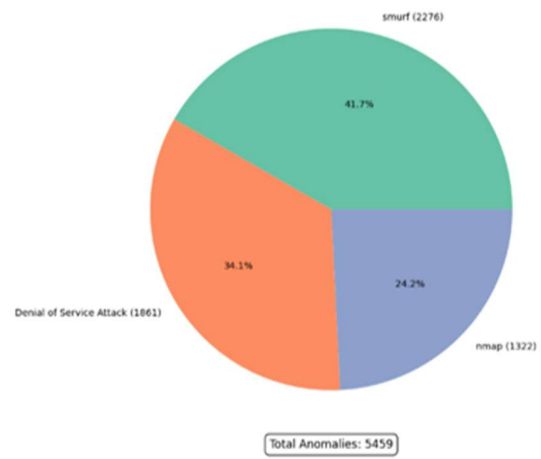

**Gambar 2.3** Fitur-fitur Yang Memiliki Angka Anomali Tertinggi Diukur Berdasarkan Fungsi Keputusan Dari *Isolation Forest*.

### 4.5. Analisis Fitur

Penerapan analisis fitur dilakukan untuk menyelidiki lebih lanjut mengapa interaksi yang dapat dianggap "normal" cenderung lebih sering terjadi. Di bawah ini, hasil eksplorasi perbedaan statistik antara kategori "normal" dan kategori lainnya untuk 10 fitur teratas dipaparkan.

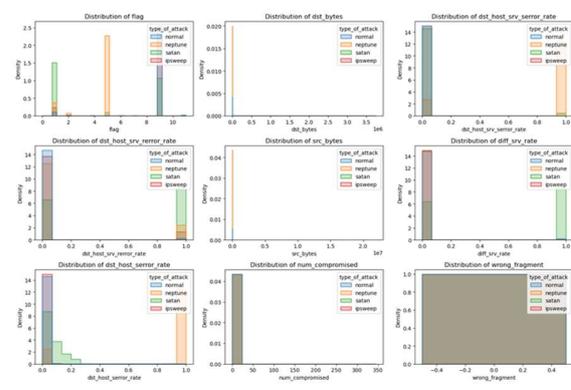

**Gambar 2.4** Persebaran Masing-masing Fitur Untuk Beberapa Kelas Dari Dataset NSL-KDD

Berdasarkan gambar di atas, dapat dilihat bahwa nilai 'flag' untuk interaksi normal berkisar antara 1 hingga 5. Selanjutnya, interaksi normal memiliki nilai 'dst_host_srv_serror_rate', 'dst_host_srv_rerror_rate', 'diff_srv_rate', dan 'dst_host_serror_rate' sekitar 0. Selain itu, interaksi normal memiliki variasi yang lebih rendah dibandingkan dengan serangan 'satan' dan 'neptune' pada 'dst_host_serror_rate'. Fitur lain seperti

'dst_bytes', 'src_bytes', 'num_compromised', dan 'wrong_fragment' sebagian besar memiliki karakteristik yang sama dengan interaksi normal.

### 4.6. Perancangan Model Generatif and Augmentasi Data

Ketidakseimbangan data diidentifikasi dalam dataset, yang berpotensi merugikan ketahanan (*robustness*) model dan akurasi klasifikasi, terutama untuk kelas-kelas serangan yang kurang terwakili. *Generative Adversarial Networks* (GAN) digunakan untuk memperbaiki ketidakseimbangan ini. Arsitektur model ditampilkan pada Gambar 2.5.

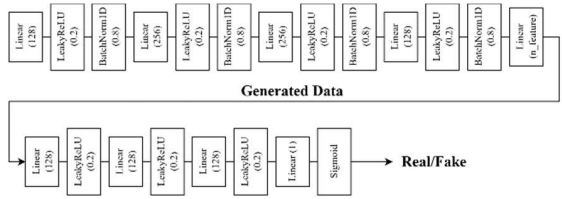

**Gambar 2.5** Arsitektur *Deep Generative Adversarial Networks* untuk Rekonstruksi Fitur Tiap Kelas Serangan

Struktur model tersusun dari empat blok generator yang masing-masing menggabungkan jaringan feed-forward, aktivasi LeakyReLU (gradien = 0.2), dan normalisasi batch, serta tiga blok diskriminator, yang masing-masing menggabungkan jaringan feed-forward dan aktivasi LeakyReLU (gradien = 0.2). Model kemudian dilatih menggunakan 5000 epoch dengan menggunakan fungsi kerugian *binary cross-entropy* yang didefinisikan sebagai berikut.

$$\mathcal{L}_{BCE} = -\frac{1}{n}\sum_{i=1}^{n}(y_i \log(\widehat{y_i}) + (1-y_i)\log(1-\widehat{y_i})), \quad (5)$$

dimana $y_i$ bernilai 1 untuk fitur data asli dan 0 untuk fitur yang dibangun oleh GAN. Tabel 4.1 dan 4.2 menggambarkan kinerja model yang dihasilkan sebelum dan setelah augmentasi.

| Class | Precision | Recall | F1-Score | Accuracy |
|---|---|---|---|---|
| **DDoS** | **1.0000** | 0.9917 | 0.9959 | |
| ipsweep | 0.9103 | 0.9707 | 0.9997 | |
| neptune | 1.0000 | 0.9997 | 0.9999 | |
| nmap | 0.9034 | 0.7480 | 0.9998 | 0.995362 |
| normal | 0.9983 | **0.9998** | **0.9991** | |
| portsweep | 1.0000 | 0.9943 | 0.9971 | |
| satan | 0.9985 | 0.9835 | 0.9909 | |
| smurf | **1.0000** | **1.0000** | **1.0000** | |

**Tabel 4.1** Performa Model Sebelum Augmentasi Data Menggunakan GAN

| Class | Precision | Recall | F1-Score | Accuracy |
|---|---|---|---|---|
| **DDoS** | 0.9996 | **0.9991** | 0.9994 | |
| ipsweep | **0.9835** | **0.9960** | 0.9897 | |
| neptune | 1.0000 | **1.0000** | 1.0000 | |
| nmap | **0.9978** | **0.9906** | 0.9942 | 0.997886 |
| normal | **0.9987** | 0.9992 | 0.9990 | |
| portsweep | 0.9999 | **0.9998** | 0.9998 | |
| satan | **0.9991** | **0.9983** | 0.9987 | |
| smurf | 0.9998 | 1.0000 | 0.9999 | |

**Tabel 4.2** Performa Model Sesudah Augmentasi Data Menggunakan GAN

Setelah memeriksa hasilnya, telah terbukti bahwa penerapan *Generative Adversarial Networks* (GAN) untuk augmentasi data secara signifikan meningkatkan keefektifan dan ketangguhan klasifikasi model. Indikator penting dari hal ini adalah peningkatan substansial dari nilai recall untuk kelas 'nmap', yang secara nyata meningkat dari 0.7480 menjadi 0.9906. Bersamaan dengan itu, akurasi model juga mengalami peningkatan dari 99.5362% menjadi 99.7886%.

## 5. Hasil Analisis
### 5.1. Presisi dan *Recall* Model
Penggunaan XGBoost menghasilkan akurasi yang mengesankan sebesar 99.54%. Namun, ketidakseimbangan kelas yang substansial, yang didominasi oleh interaksi 'normal' dan diikuti oleh berbagai tingkat serangan 'neptune, 'satan', dan 'ipsweep', mengindikasikan potensi risiko bias model terhadap kelas-kelas yang terlalu banyak diwakili. Selain itu, meskipun augmentasi GAN secara khusus meningkatkan akurasi model menjadi 99.78% dan recall untuk kelas 'nmap' dari 0.7480 menjadi 0.9906 dan meningkatkan akurasi secara keseluruhan, ketepatan, terutama untuk kelas yang kurang terwakili, memerlukan pemeriksaan lebih lanjut untuk memastikan model tidak menghasilkan positif palsu secara berlebihan.

### 5.2. Tingkat Kepentingan Fitur and Interaksi

Identifikasi fitur-fitur penting seperti 'flag', 'dst_bytes', dan 'dst_host_srv_serror_rate' memberikan wawasan yang berharga ke dalam variabel-variabel yang paling berpengaruh dalam mengklasifikasikan interaksi dan serangan. Namun, memahami interaksi antar fitur, terutama dengan mempertimbangkan berbagai perilaku yang dicatat antara jenis 'normal' dan serangan dalam fitur seperti 'dst_host_serror_rate', dapat menyingkap hubungan dan ketergantungan yang mungkin dapat menyempurnakan kemampuan klasifikasi.

### 5.3. Identifikasi Anomali dan *Attack Vector* Langka

Penggunaan algoritma Isolation Forest dalam analisis anomali menunjukkan potensi untuk mengidentifikasi dan memahami vektor serangan yang jarang terjadi. Namun, memahami tingkat positif palsu, dan memastikan bahwa anomali benar-benar menunjukkan serangan yang jarang terjadi dan bukannya salah klasifikasi, sangat penting untuk menjaga kredibilitas deteksi anomali dalam aplikasi keamanan siber.

Kesimpulannya, paper ini telah melakukan analisis eksploratif untuk menggali *insight* lebih dalam mengenai dataset NSL-KDD dan menunjukkan adanya ketidakseimbangan distribusi data berdasarkan kelas. Masalah ini diatasi dengan menggunakan GAN untuk melakukan augmentasi data dengan mempelajari fitur-fitur yang ada dari data latih. Sebagai hasilnya, metode ini memberikan hasil yang signifikan jika dibandingkan dengan penelitian-penelitian sebelumnya.

## Referensi

# LAMPIRAN

**1. Grafik *Loss* Pelatihan DeepGAN Untuk Setiap Data Kelas Target (dari atas ke bawah: nmap, denial of service, smurf, portsweep, ipsweep, dan satan)**

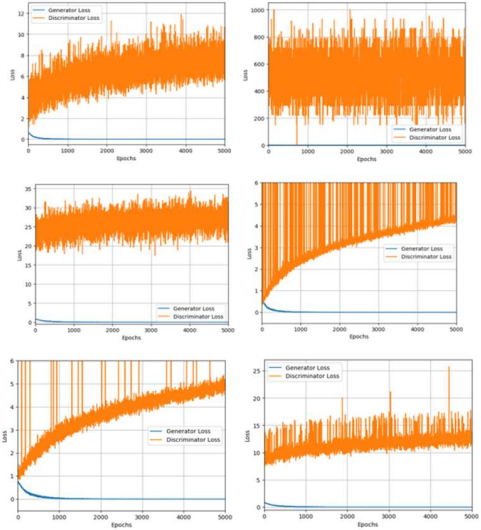

**2. Confusion Matrix Klasifikasi Model (Sebelum dan Sesudah Augmentasi Data)**

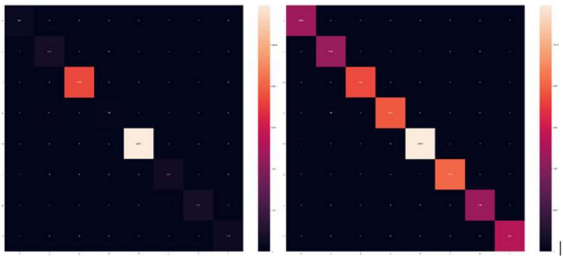

**3. Distribusi Data Yang Direkonstruksi Dan Data Asli untuk Kelas Serangan 'nmap'**

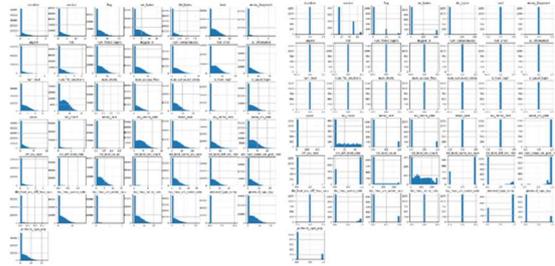

**4. Perbandingan Durasi Koneksi Untuk Setiap Kelas**

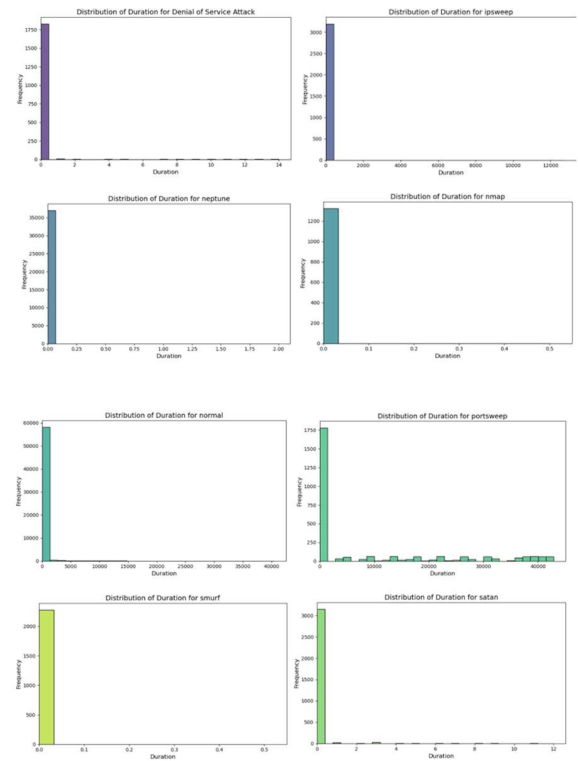